\documentclass[aps,pre, preprint, onecolumn]{revtex4}
\usepackage{amssymb}
\usepackage{eurosym}
\usepackage{epsfig}

\begin{document}

\normalsize

\title{Dislocations of soliton lattices: experiment and theory}
\author{E.A. Kuznetsov $^{(a,b)}$} 
\affiliation{$^{(a)}$ P.N. Lebedev Physical Institute of RAS, Moscow, Russia\\
$^{(b)}$ L.D. Landau Institute for Theoretical Physics of RAS, Chernogolovka, Moscow region, Russia}

\begin{abstract}
The results of recent experiments \cite{Mao} on observing soliton lattices and their dislocations in vertical cylindrical channels filled with immiscible fluids with strongly different viscosities and but slightly different densities are discussed. The less viscous, lower-density fluid fills the central region of the cylinder. Injecting a light fluid from below generates nonlinear cnoidal waves at the interface between the fluids, which have the appearance of soliton lattices. Two types of lattice dislocations are observed, the interaction between which is elastic. This experimental study fully confirms the theory of cnoidal waves and their dislocations for the KDV equation, which was developed 50 years ago and published in JETP \cite{KuznetsovMikhailov}.                            

\end{abstract}

\maketitle
\tableofcontents

\section{Introduction}
This paper is motivated by a recent (2023) experimental study \cite{Mao}, which observed dislocations propagating along a periodic cnoidal wave. The authors call them traveling breathers. These large-amplitude breathers oscillate in time during their propagation, but on average, they move at a constant velocity. After passing a breather, the cnoidal wave recovers, acquiring a spatial shift relative to the original cnoidal wave; i.e., these breathers represent moving dislocations of the cnoidal wave. The experiments revealed two types of breathers—dark and bright—which move faster or slower than the cnoidal wave. It was also experimentally shown that the collision of breathers-dislocations is elastic.
All these facts confirm the theoretical predictions developed in the work of the author and A.V. Mikhailov \cite{KuznetsovMikhailov}, published 50 years ago.
In this paper, we will examine in detail the results of the  experiment \cite{Mao} and, following \cite{KuznetsovMikhailov}, provide corresponding theoretical explanations for the experiments.

\subsection{Background and Experiment}
The main focus of experiments \cite{Mao} is the study of nonlinear wave structures that arise at the interface of two liquids with very different viscosities in a tall vertical column with dimensions ($5 cm \times 5 cm \times 180 cm $). We begin with a description of the experiments, which is largely outlined in the work \cite{MaoHoefer},  preceded \cite{Mao}. One liquid, heavier, with a significantly higher viscosity, occupies the outer vertical region bounded by the wall, while the other region—the interior of the cylindrical column—is filled with a different liquid with lower viscosity and lower density. In the experiments \cite{MaoHoefer, Mao}, glycerol was used as the heavier but more viscous liquid, and the internal liquid was an aqueous glycerol solution with a density lower than the external liquid. The viscosity ratio was $\epsilon=0.04$. The region of less viscous fluid, immiscible with the surrounding fluid, in a steady state was a cylinder with a radius constant along the column. Obviously, if a localized disturbance of larger radius were created somewhere within this cylinder, this disturbance would begin to flow upward due to convective instability. This disturbance, in the form of a soliton, would propagate upward along the column at a constant velocity.

This fact was first established numerically and analytically, and also confirmed experimentally, in the works \cite{ScottStevensonWhitehead,  OlsonChristensen} (see also \cite{HelfrichWhitehead}). As was clarified in these works, with a local change in the radius of the inner cylinder, the soliton was a localized disturbance propagating upward at a constant velocity while maintaining its shape. It is noteworthy that in 1984, Scott and Stevenson proposed using this model to explain the motion of a magma clot, called a magma soliton, which propagates upward along a volcanic conduit during an eruption. Magma in the conduit has a lower viscosity than the conduit walls, which was the main motivation for these authors.

In the experiments \cite{MaoHoefer, Mao} both linear and nonlinear waves, including solitons, were generated by injecting a light, less viscous liquid from below. A nonlinear model for the propagation of such waves, called the conduit equation, was developed 
\cite{OlsonChristensen, LowmanHoefer}. This equation is written in dimensionless variables for the area $A$ of the inner cylinder (normalized to the unperturbed value), which depends on the vertical coordinate $z$ and time $t$:
\begin{equation} \label{conduit}
A_t + (A^2)_z - (A^2(A^{-1}A_t)_z)_z = 0.
\end{equation}
This equation was initially derived for stationary waves 
\cite{OlsonChristensen}, and then its unsteady version (\ref{conduit}) was obtained in \cite{LowmanHoefer}. The derivation used a small parameter — the ratio of the viscosities of the internal and external fluids — and the condition of  forces equality at the interface, assuming a small Reynolds number of about unity. This means that the internal fluid flow in each cross-section can be considered a Poiseuille flow due to the vertical pressure gradient caused by buoyancy (the internal fluid has a lower density than the external fluid). It is interesting to note that the conduit equation is formally conservative, despite the presence of viscous forces. A similar situation arises when describing the motion of the interface between two Hele-Shaw flows with very different viscosities. As is well known, such boundary motion is described by the so-called Laplace growth equation, first obtained in the works 
\cite{kochina, galin}. Within this equation, the famous Saffman-Taylor solution was obtained, shaped like a finger penetrating into a more viscous medium.

The conduit equation (\ref{conduit}) in the linear approximation, when $A=1+\alpha$ ($|\alpha|\ll 1$), describes small-amplitude waves with the dispersion law
\begin{equation} \label{disp}
\omega=\frac{2k}{1+k^2}.
\end{equation}
In the experiments of \cite{MaoHoefer, Mao} for cnoidal waves and breathers propagating along them, the characteristic value of the dimensionless wave number $k$ was of the order of $0.3$ - $0.4$, so the dispersion additive was small: $k^2 \approx 0.1$. Thus, for the experiments of \cite{MaoHoefer, Mao}, the dispersion law is close to the dispersion law for the KDV equation: $ \omega =2k(1-k^{2})$.
In a nonlinear regime, for example, for a stationary cnoidal wave, this means that dispersion must be balanced by nonlinearity. Therefore, for the experimental parameters
\cite{MaoHoefer, Mao}, in a nonlinear regime, waves should be well approximated by the KDV equation:
\begin{equation} \label{KDV}
u_t + u_{xxx} + 6 uu_x=0.
\end{equation}
This equation is written in standard form and is obtained from (\ref{conduit}) using simple transformations: transition to a moving coordinate system and appropriate scaling (assuming weak dispersion and low nonlinearity). The latter, as will be shown below, allows us to explain all the experimental results of the paper  \cite{Mao}.

\subsection{Experimental datas}

Let's turn to the experimental results \cite{Mao}. Figure 1a shows observations at different moments of time of a bright breather propagating along a cnoidal wave. The bottom panel corresponds to the unperturbed cnoidal wave. The upper panels show that the breather moves faster than the cnoidal wave. Figure 1b shows the breather's trajectory in the $t-z$plane: it is the dashed red line against a background of straight blue lines corresponding to the cnoidal wave's motion. As can be seen, the breather moves at a constant speed on average, undergoing small oscillations. This figure also shows a slight shift in the cnoidal wave after the breather's passage.
\begin{figure}[tbp]
\centering
\includegraphics[width=7.5cm]{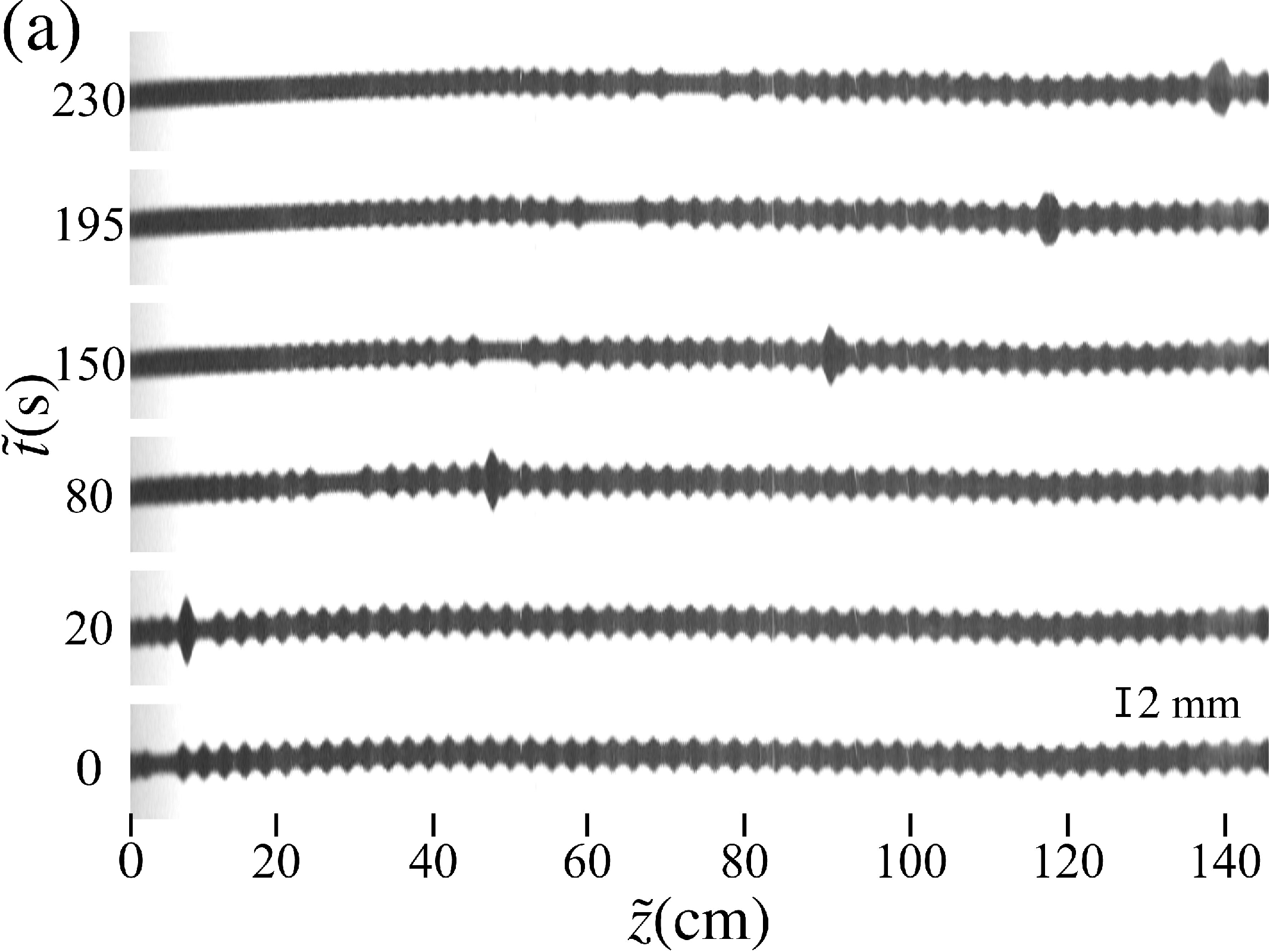}
\includegraphics[width=7.5cm]{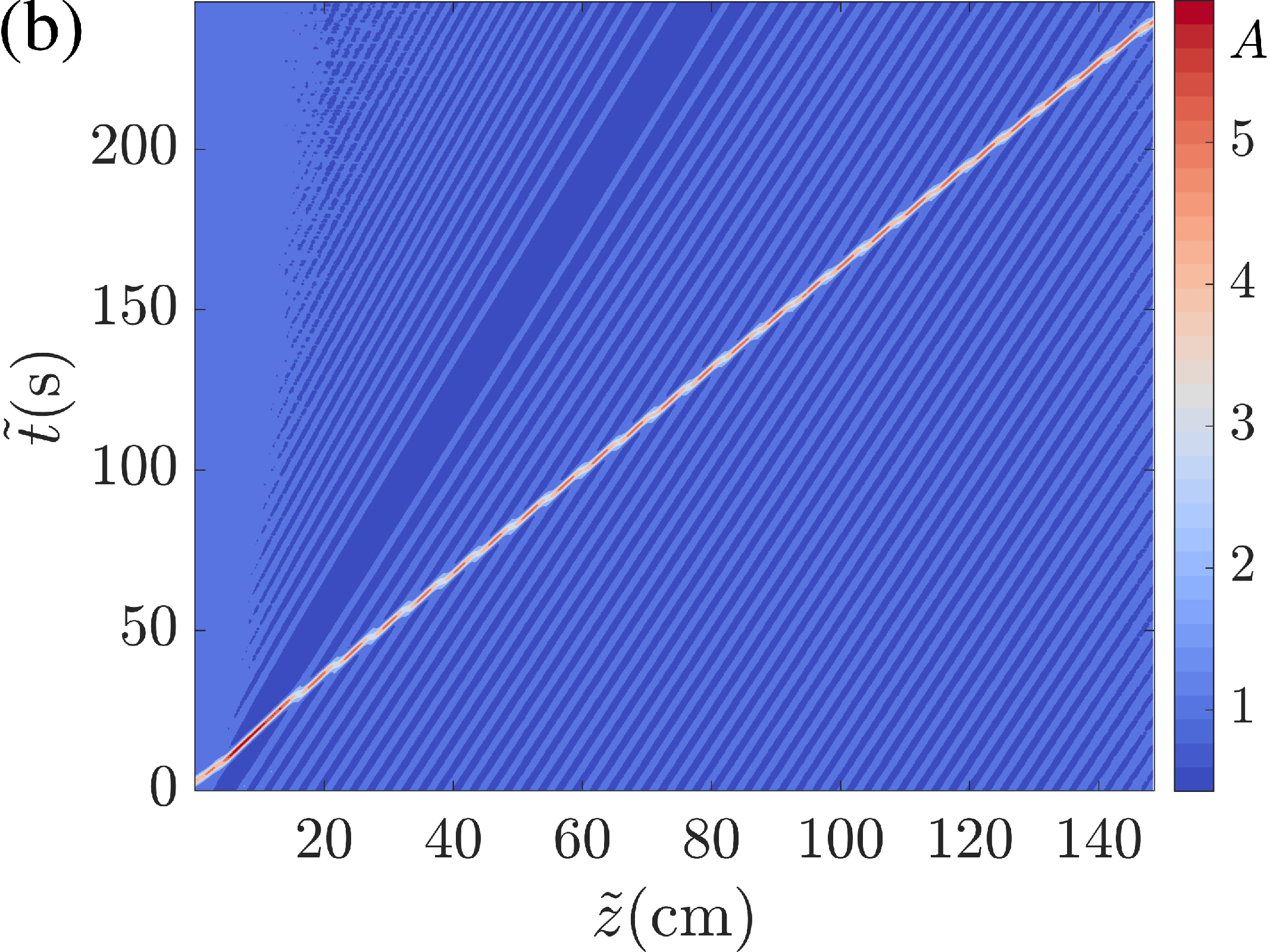}
\caption{
(a)The motion of a bright breather along a cnoidal wave at different moments in time.
The bottom panel corresponds to the unperturbed cnoidal wave.
(b) The trajectory of a bright breather on the t-z plane (dashed red line) against a background of straight blue lines corresponding to the propagation of the cnoidal wave.
}
\end{figure}

Figure 2b shows the trajectory of a dark breather: its average velocity is lower than the cnoidal wave velocity, and the breather oscillates. After its passage, it forms a cnoidal wave shift of the opposite sign compared to the bright breather.

\begin{figure}[tbp]
\centering
\includegraphics[width=7.5cm]{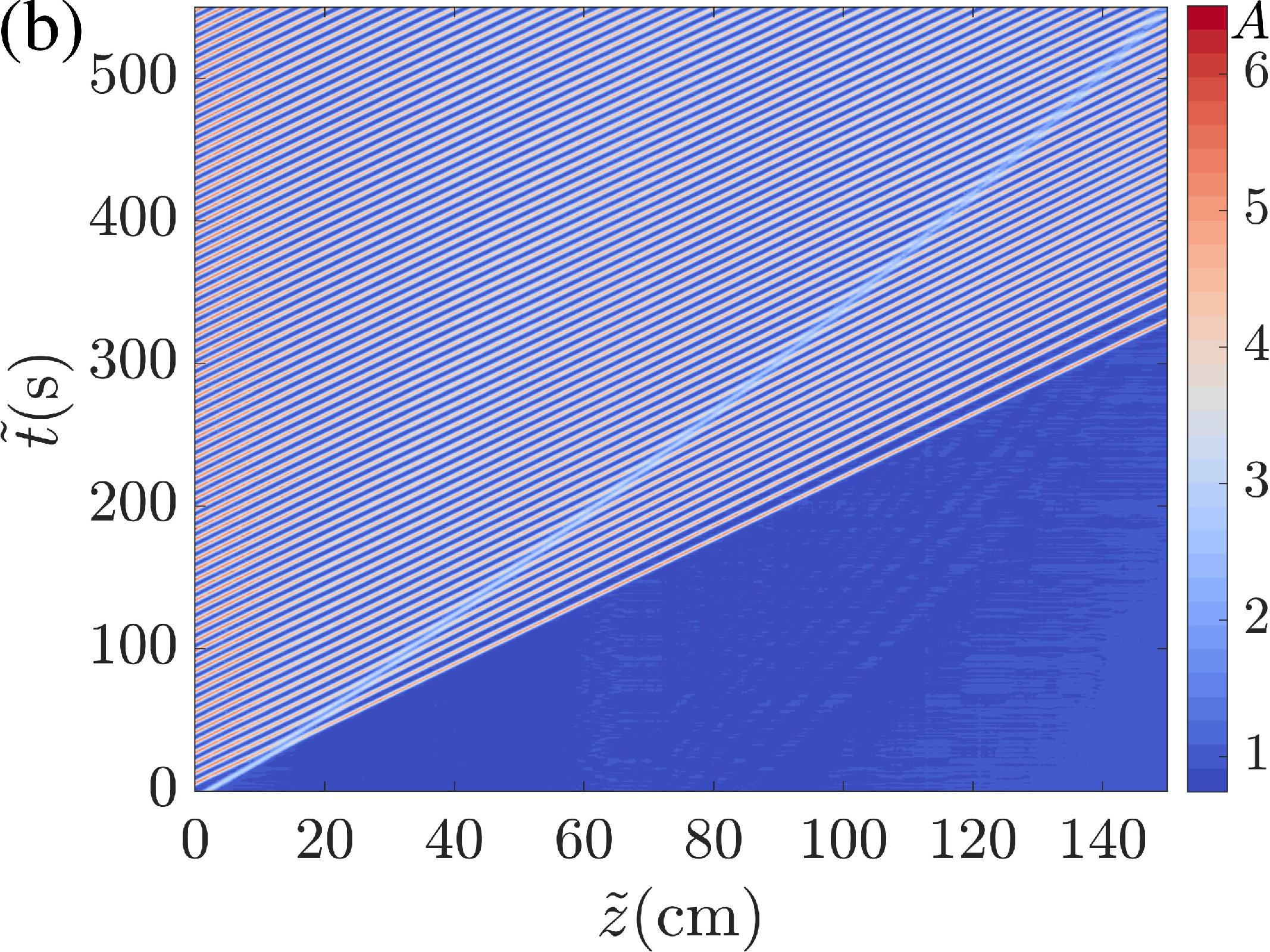}
\caption{
Dark breather trajectory on the $t-z$ plane (light blue line) against the background of the cnoidal wave.
}
\end{figure}
Thus, the interaction of breathers with a cnoidal wave does not change its velocity or, importantly, its amplitude, but leads to the formation of dislocations. As will be shown below, this is a consequence of the integrability of the KDV equation, which describes the behavior of bright and dark breathers in this experiment with good accuracy.

A very interesting result of these experiments was also the study of breather interactions, which showed that the scattering of both bright and dark breathers between themselves is elastic: Figure 4a shows the scattering of bright breathers, Figure 4c the scattering of dark breathers, and Figure 4e the scattering between a bright breather and a dark breather.
\begin{figure}[tbp]
\centering
\includegraphics[width=5.0cm]{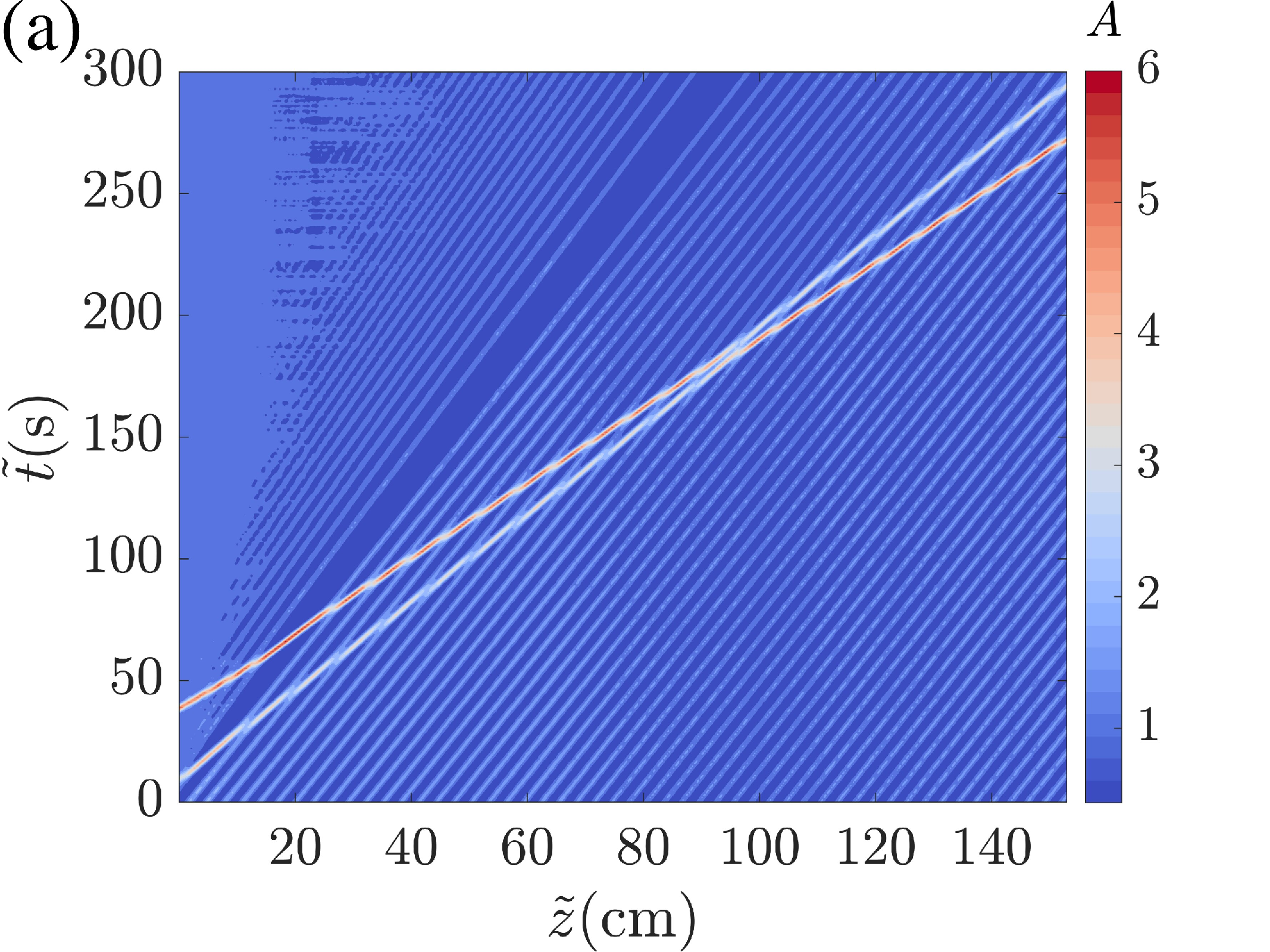}
\includegraphics[width=5.0cm]{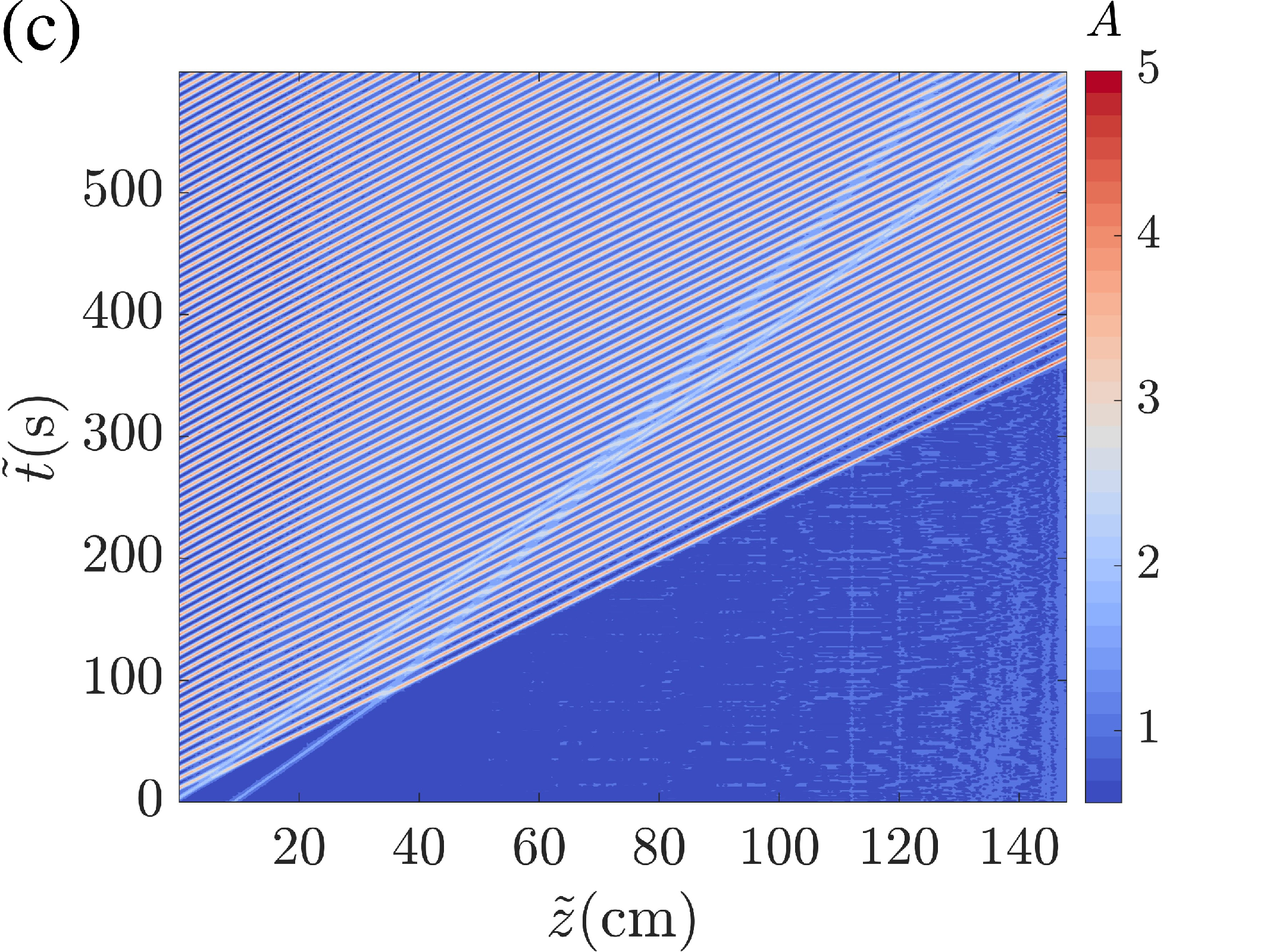}
\includegraphics[width=5.0cm]{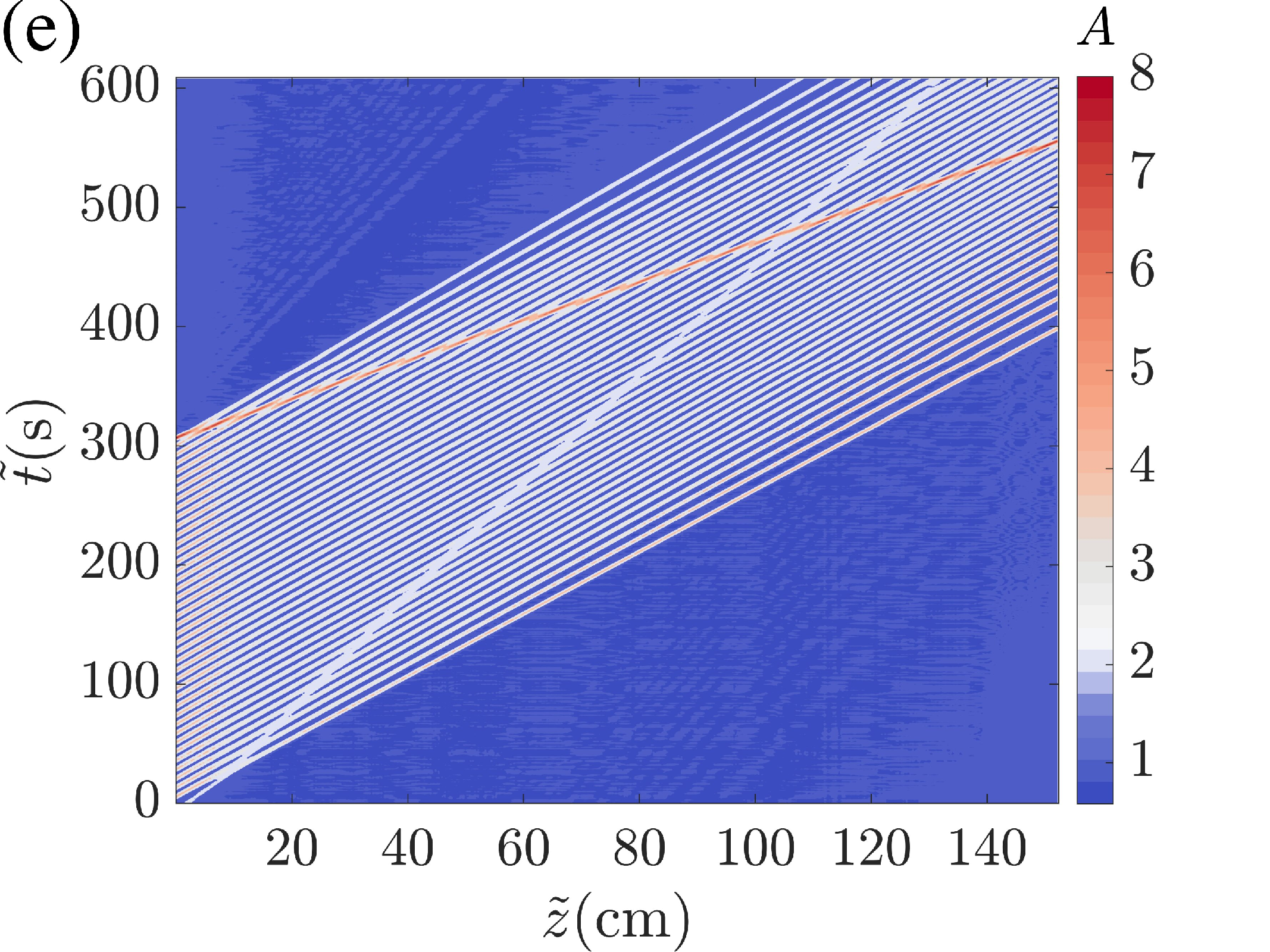}
\caption{
(a) Scattering of two bright breathers.
(c) Scattering of two dark breathers.
(e) Scattering of a bright and dark breather.
}
\end{figure}
 With all these scatterings, the (average) velocities of the breathers remain constant. As will be shown in the next section, these results also confirm the theoretical predictions developed in the work \cite{KuznetsovMikhailov}.

\section{Theory of cnoidal waves and their defects}

In study of the stability of the KDV cnoidal wave \cite{KuznetsovMikhailov} , we first applied a dressing procedure, the Shabat scheme \cite{Shabat}, which is based on the Marchenko integral equation by introducing an algebra of differential operators with respect to the convolution of two kernels.
Later, Zakharov and Shabat \cite{ZakharovShabat} presented a more developed version of the dressing method, based on the Riemann-Hilbert problem.
This scheme is now called the Zakharov-Shabat dressing procedure.

The dressing procedure for systems integrable using the inverse scattering method allows one to construct a new solution based on a known solution.
In this formulation, this method is similar to the well-known Darboux transformation, but not completely. The dressing method proved to be more effective than the Darboux method. It allows ones to study the nonlinear dynamics of solutions with nontrivial asymptotic behavior at infinity, in particular, to investigate the Cauchy problem for such boundary conditions. In this sense, Shabat's scheme proved more productive than the Darboux method. Specifically, using this scheme, we solved the stability problem of a cnoidal wave of the KDV. We investigated not only linear but also nonlinear stability.
It should be noted that the dressing method offers significant advantages for analyzing the stability of an arbitrary solution to integrable equations compared to the standard procedure based on direct linearization.
In a series of papers \cite{Kuznetso1977, KuznetsovSpectorFalkovich, KuznetsovSpector, Kuznetsov-2017}, this approach was applied to both linear and nonlinear stability problems of cnoidal waves for the Kadomtsev-Petviashvili equation and the nonlinear Schrödinger equation.

\subsection{Shabat scheme}

Consider the Marchenko equation for two kernels $K(x,y)$ and $F(x,y)$
\begin{equation} \label{marchenko}
K(x,y)+F(x,y)+\int_x^{\infty} K(x,s)F(s,y)ds=0
\end{equation}
and introduce the differentiation operators $D$ with respect to the convolution of the two kernels
\[
G*H=\int_{-\infty}^{\infty} G(x,s)H(s,y)ds: \,\, D(G*H)=(DG*H)+(G*DH).
\]
The simplest differential operators are easily established:
\[
D_t=\frac{\partial}{\partial t}, \,\, D_n=\frac{\partial ^n}{\partial x^n}+(-1)^{n+1}\frac{\partial ^n}{\partial y^n},\,\, D_f=f(x,t) - f(y,t),
\]
which form a basis for the algebra: the commutator of two operators and their linear combination are also differentiation operators.

According to the Shabat' theorem \cite{Shabat}, if the kernels $K$ and $F$ obey the Marchenko equation
(\ref{marchenko}), and furthermore $DF=0$, where $D$ is any operator in the algebra, then there exists an operator $\tilde{D}$ such that $\tilde{D}K=0$. This fact is verified by applying the operator $D$ to (\ref{marchenko}) and integrating by parts.

We will show how the theorem works for the KDV equation (\ref{KDV}), which
is written in the Lax representation:
\begin{equation} \label{lax}
\frac{\partial L}{\partial t}=[L,A],
\end{equation}
where
\[
L(x)=\frac{\partial ^2}{\partial x^2}+u,\,\,
A(x)= 4\frac{\partial ^3}{\partial x^3}+ 6u\frac{\partial}{\partial x}+ 3u_x,
\]
and $u$ is a solution of equation (\ref{KDV}).

Let $F$ satisfy two equations
\begin{equation} \label{bare}
PF=(L(x)-L(y))F=0,\,\, QF=(\partial_t+A(x)+A(y))F=0,
\end{equation}
which are compatible by virtue of the representation (\ref{lax}) for some solution (\ref{KDV}) $u=u_0$. Here $L(x)$ means that the operator $L$ depends on $x$ and its derivatives with respect to $x$, and $L(y)$ depends on $y$ and its derivatives with respect to $y$, etc. It is easy to verify that the introduced operators $P$ and $Q$ are differentiation operators.

The action of $P$ and $Q$ on the Marchenko equation yields the following two differential equations for the kernel $K$:
\begin{equation} \label{dressing}
{\tilde P}K=0,\,\, {\tilde Q}K=0.
\end{equation}
Here $ {\tilde P}=P+2\frac{d}{dx}K(x,x)$ is the result of dressing.
The compatibility of these two equations leads to the fact that
$u(x)=u_0(x)+2\frac{d}{dx}K(x,x)$ satisfies the KDV equation, where $2\frac{d}{dx}K(x,x)=w$ is the perturbation of $u_0(x)$, vanishing as $x\to \infty$.
Thus, this method allows one to effectively study the stability of any solutions with nontrivial asymptotics with respect to any perturbations, including finite ones.

Note that both equations (\ref{bare}) and (\ref{dressing}) allow separation of variables with respect to $x$ and $y$. In particular, after separation of variables, the resulting equations are consistent due to the Lax representation (\ref{lax}).

\subsection{Cnoidal wave as a soliton lattice}

A cnoidal wave is a stationary periodic solution
to the KDV equation, moving with velocity $v$: $u(x-vt)=2 \wp(x+i\omega'-vt)+v/6$
where $\wp(x)$ is the elliptic doubly periodic Weierstrass function with periods $2\omega,\,2i\omega'$.
Since the KDV equation is Galilean invariant, we can set $v=0$.
In this case, a partial solution for $F$ can be written as
\begin{equation} \label{F}
F(x,y,t)=C(t)\psi(x)\psi(y),
\end{equation}
where $\psi$ satisfies the Schrödinger equation
with potential $u$:
\begin{equation} \label{sch}
[d^2/dx^2 -2\wp(x+i\omega')]\psi=-E\psi.
\end{equation}
This equation is known as the Lamé equation (see, e.g., \cite{WhittakerWatson}).
Its eigenfunctions are expressed in terms of the Weierstrass functions $\sigma(x)$ and
$\zeta (x)$:
\begin{equation} \label{eigen}
\psi_a(x)= \frac{\sigma (x+i\omega'+a)}{\sigma (x+i\omega')\sigma (a)}\exp[\zeta(a)x +\zeta(i\omega')a],
\end{equation}
where the energy $E=-\wp(a)$. The functions $\psi_a$ and $\psi_{-a}$ are linearly independent. In this case, the parameter $a$ plays the role of the spectral parameter. From the properties of the Weierstrass function it follows that
$E$ takes real values in the complex plane of the parameter $a$ on four segments (see Fig. 4): two segments $(\omega, \omega+i\omega'),\, (i\omega', i0)$ correspond to the continuous spectrum (two allowed bands):
$-\wp(\omega)\geq E \geq -\wp(\omega+i\omega')$ and
$-\wp(i\omega')\geq E \geq -\wp(i0))=\infty$.
The other two segments $(i\omega', \omega+i\omega'),\, (\omega, 0)$ define the forbidden bands (with $E(0)=-\infty$!).
\begin{figure}[tbp]
\centering
\includegraphics[angle=0,width=8cm]{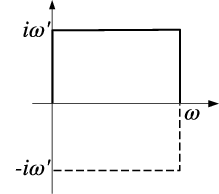}
\caption{Band positions in the complex plane $a$.}
\end{figure}
The wave function (\ref{eigen}), corresponding to the continuous spectrum, has a Bloch form with quasi-momentum $p(a)=i\omega^{-1}(\omega\zeta(a))-\zeta(\omega)a)$. Moreover, $p(a)$ is continuous when passing from one continuous band to another.

According to S.P. Novikov's definition \cite{Novikov}, the solution in the form of the Weierstrass function $u=-2\wp(x+i\omega')$ belongs to the so-called single-zone potential (see also \cite{ZakharovManakovNovikovPitaevsky}).

As shown in \cite{KuznetsovMikhailov}, this potential can be represented as a periodic soliton lattice:
\begin{equation} \label{lattice}
u(x)=2\left( \frac{\zeta(i\omega')}{i\omega'}
+\sum_{n=-\infty}^{\infty}\frac{\kappa^2}{\mbox{ch}^2\kappa(x+2n\omega)}\right)
\end{equation}
where $\kappa=\pi/(2\omega')$.
Each term of this sum represents a soliton
\[
u_s=\frac{2\kappa^2}{\mbox{ch}^2\kappa(x)},
\]
which as a solution is obtained in the limit of a large period $\omega\to\infty$. This soliton potential is known to be reflectionless. $u_s$ has only one bound state with $E_s=-\kappa^2$. From the standpoint of quantum mechanics, it becomes qualitatively clear why the potential (\ref{lattice}) is single-band. If we take two solitons $u_s$, separated from each other by a large distance, then the $E_s$ level will split into two. In the case of a periodic lattice of solitons, an allowed band will appear near $E_s=-\kappa^2$ at these energies. The question arises as to why there are no forbidden bands at high energies, $ E \to +\infty$. The reason is related to the reflectionless nature of the single-soliton potential $u_s$. If at high energies the potential $u_s$ had an exponentially small reflection coefficient, then a forbidden band would appear at these energies. Due to the absence of reflection, there are no such bands. This is a qualitative explanation that requires a more precise analysis. This raises a natural question: wouldn't a two-band potential arise if, for example, we take a two-soliton potential instead of $u_s$ in (\ref{lattice})? This is obviously correct if we consider $u=-6\wp(x+i\omega')$ instead of $u=-2\wp(x+i\omega')$.

Substituting (\ref{F}) into (\ref{bare}) yields the dependence $C(t)$:
\[
C(t)=C(0)\exp[-4\wp'(a)t]
\]
Hence, the general expression for $F$ will be defined as follows:
\begin{equation} \label{F-gen}
F(x,y,t)=\int_c\rho(a,t)\psi_a(x)\psi_a(x)da +\sum_n M^2_n(t) \psi_n(x)\psi_n^*(y),
\end{equation}
where the integration corresponds to the continuous spectrum, and the discrete sum, as usual, determines the contribution from "solitons" for which $Im \,p(a_n)>0$. The latter ensures that $F$ vanishes as $y\to\infty$. The temporal behavior of the coefficients $\rho(a,t)$ and $M^2_n(t)$ is determined by the temporal dependence $C(t)$.

Note that representing a cnoidal wave as a soliton lattice allows us to understand many aspects of the interaction of the lattice with an off-lattice soliton propagating along the lattice.

\subsection{Soliton as a dislocation of the soliton lattice}

Substituting (\ref{F-gen}) into (\ref{marchenko}) yields the general solution for the kernel $K(x,y)$ in the form
\[
K(x,y,t)=-\int_c\rho(a,t)\phi_a(x)\psi_a(x)da -\sum_n M^2_n(t) \phi_n(x)\psi*_n(y),
\]
where the functions $\phi_a(x)$ and $\phi_n(x)$ are found from the triangular representation,
\[
\phi_{a, n}(x)=\psi_{a,n}(x) +\int_x^{\infty} K(x,y)\psi_{a,n}(y) dy.
\]
Both sets of these functions satisfy the Schrödinger equation (\ref{sch})
\[
[d^2/dx^2 + u(x)]\phi=-\wp(a)\phi,
\]
where $u=u_0+w$. The function $\phi_a(x)$ corresponds to the continuous spectrum, $\phi_n(x)$ to the discrete spectrum of the potential $u(x)$.

We consider the simplest case, when there is no contribution from the continuous spectrum, and we retain only one term in the discrete sum, which corresponds to the new one-soliton solution. Two options are possible: either $a$ lies on the interval $(\omega, 0)$, i.e., corresponds to the lower forbidden band, or on the interval
$(i\omega', \omega+i\omega')$, corresponding to the upper forbidden band. The spatial dimensions of the resulting solutions will obviously be different. In the first case, the size will be smaller than the size of a soliton from the lattice $\sim\kappa^{-1}$, and the amplitude will be correspondingly greater than $2\kappa^{2}$, which obviously follows if we consider the limit of a large period $\omega\to\infty$. In the second case, everything will be the opposite. This conclusion follows directly from the solution of the Marchenko equation:
\[
u(x,t)= u_0(x) +2 \frac{d^2}{dx^2}\mbox{ln}\left(1+M_n^2(t)\int_x^{\infty}|\psi_n|^2dy\right).
\]
The integral here is calculated explicitly:
\[
\int_x^{\infty}|\psi_n|^2dy=|\psi_n|^2\frac{\sigma(x'+2 \mbox{Re}\,a)\sigma(x')|\sigma(a)|^2}{\sigma(2 \mbox{Re}\,a)\sigma(x'+a)\sigma(x'+a*)}
\]
where $x'=x+i\omega'$.

Analysis of this solution shows (see \cite{KuznetsovMikhailov} ) that $u(x,t)\to u(x+2 \mbox{Re}\,a)$ as $x\to -\infty$. In other words, we have a lattice reconstruction that shifts by $2 \mbox{Re}\,a$ relative to the lattice at the other end, as $x\to \infty$.
Thus, the one-soliton solution represents a dislocation of the soliton lattice. This dislocation is a non-stationary oscillating object, its average velocity being $V=-2\wp'(a)/\mbox{Im}\,p(a)$. All characteristics of this dislocation are determined by the spectral parameter $a$, with the exception of the spatial position of the dislocation itself. Depending on the position of the parameter $a$, whether it lies in the first or second forbidden band, the sign of the velocity $V$ is determined; i.e., the dislocation moves faster or slower than the cnoidal wave, which is in complete agreement with the  experiment \cite{Mao}. Fast dislocations correspond to bright breathers, and slow dislocations to dark breathers.
Figure 5, taken from the paper \cite{KuznetsovMikhailov}, shows solutions in the form of these two types of dislocations. The dependence of the cnoidal wave on $x$ is shown for comparison.

\begin{figure}[tbp]
\centering
\includegraphics[angle=0,width=10cm]{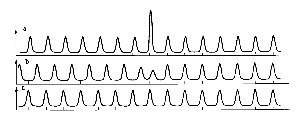}
\caption{ (a) Soliton in the form of a dislocation with parameters from the first band gap. (b) Soliton in the form of a dislocation with parameters from the second band gap. (c) Cnoidal wave.}
\end{figure}

The interaction of a cnoidal wave with an off-lattice soliton propagating along the lattice, leading to the appearance of a dislocation, has a very simple explanation. As is known, the interaction between KDV solitons is elastic and pairwise, which, upon scattering, only leads to a shift in their centers of mass \cite{GardnerGreeneKruskalMiura} (see also
\cite{ZakharovManakovNovikovPitaevsky}).
Consequently, a soliton propagating along the lattice, after scattering from a lattice soliton, will restore that lattice soliton, which will only acquire a certain spatial shift. This will also happen with each lattice soliton colliding with an off-lattice soliton. It should be noted that all lattice solitons are identical. Thus, the soliton lattice will restore itself, acquiring a spatial shift. The soliton propagating along the lattice, as a lattice defect, senses not only the nearest lattice soliton but also those following it. As a result, the soliton, as a lattice dislocation, will move only at a constant velocity on average, undergoing small oscillations. Note that such restoration of the soliton lattice can be considered an analogue of the Fermi-Pasta-Ulam recurrence phenomenon.

As noted above, there are two types of dislocations. The first corresponds to a spectral parameter lying in the first band gap, while the second corresponds to this parameter belonging to the second band gap.
In the first case, the soliton-dislocation amplitude will be greater than the lattice soliton amplitude. Consequently, due to the properties of KDV solitons, the first type of dislocation will move faster than the cnoidal wave. In the second case, the dislocation amplitude will be less than the lattice soliton amplitude. In the first case, we have a bright breather, and in the second, a dark breather.

It is also possible to obtain a solution in the form of N dislocation solitons moving along the lattice \cite{KuznetsovMikhailov}. Scattering between dislocations, as well as between KDV solitons, will be elastic, since each dislocation is determined by its own spectral parameter $a$, which does not change over time. Only the dislocation centers change. This is precisely the behavior observed during the scattering of bright and dark breathers between themselves in the Mao experiment (see Fig. 3).
Therefore, asymptotically, as $t \to \infty$, we obtain the same cnoidal wave (shifted due to the interaction). Note that the perturbation from continuous spectra also cannot destroy the cnoidal wave, which only leads to some additional shift.

It should be noted that the response in the form of a dislocation on a single-band potential was obtained after our work by I.M. Krichever \cite{Krichever}, based on a reduction of the finite-zone integration method developed by S.P. Novikov \cite{Novikov} for integrable equations, in particular for the KDV equation.

\section{Concluding Remarks}
Thus,  the experiments \cite{Mao} fully confirmed the theoretical predictions of work \cite{KuznetsovMikhailov}: this applies to two types of breathers, bright and dark, which move at a constant average velocity, with some oscillations. Each type of breather represents a dislocation of a cnoidal wave as a soliton lattice. Bright breathers move faster than the cnoidal wave, while dark breathers move more slowly. The difference between the two dislocation-breathers is related to the spectral problem: bright breathers are determined by a spectral parameter from the lower band gap, while dark breathers are determined by a spectral parameter from the upper band gap for the integrating operator—the Schrödinger operator. Since the spectral parameters of breather-dislocations remain constant during their interaction, their scattering is elastic, with only their spatial centers changing.
It can be argued that a cnoidal wave, as a periodic lattice of solitons, is stable with respect to any perturbations, not just small ones.

That is, we can say that a cnoidal wave, as a soliton lattice, has the same degree of universality as an individual KDV soliton. In fact, in this case, we have an analogue of the Fermi-Pasta-Ulam return: under any localized perturbations, the soliton lattice returns to its initial state, experiencing a certain spatial shift as a whole. A similar situation arises when studying the nonlinear stage of modulation instability of cnoidal waves in the focusing nonlinear Schrödinger equation (NSE)
\[
i\psi _{t}+\psi _{xx}+2|\psi |^{2}\psi =0.
\]
The simplest solution in the form of a cnoidal wave is obtained by seeking the NLS solution in the form $\psi =e^{i\lambda ^{2}t}\psi _{0}(x)$, assuming
$\psi _{0}(x)$ to be a real function. Then, the equation for
the intensity $I=\psi _{0}^{2}$, after
shifting by ${\lambda ^{2}/3}$, becomes an equation for the elliptic Weierstrass function, for which, as demonstrated above, a representation in the form of a soliton lattice (\ref{lattice}) is valid.
Due to the integrability of the NLS \cite{ZakharovShabatNLS}, soliton scattering, just like in the QDV, is elastic, in which not only the coordinates of the soliton mass centers but also their phases change. However, there is one important difference between the cnoidal KDV wave and the cnoidal wave in the NSE. This is evident, for example, if the lattice period in the KDV equation tends to zero, when $\wp\to\mbox{const}$, this leads to a constant velocity, which is obviously excluded upon transition to the corresponding coordinate system. In the NSE, such a transition is impossible: the corresponding solution is a condensate, which is modulation unstable. The cnoidal wave solution \cite{KuznetsovSpector} also turns out to be unstable with respect to small modulation perturbations. In the nonlinear regime, however, a return to the unstable cnoidal wave solution occurs, which additionally receives a spatial shift and a phase shift. That is, a Fermi-Pasta-Ulam type recurrence \cite{Kuznetsov-2017} occurs. This is associated with the same phenomenon as in the KDV. In the NLS, a soliton propagating along a cnoidal soliton lattice also restores it, leading to an additional spatial shift of the lattice and its phase rotation as well. In this regard, the solution in the form of a breather propagating along a constant condensate is quite indicative. This solution was first constructed by Tajiri and Watanabe \cite{TajiriWatanabe} and then studied in detail in \cite{ZakharovGelash}. This breather moves along the condensate, leaving behind a condensate with a rotated phase but a constant amplitude. Since the condensate is an ultra-dense soliton lattice ($\omega\to 0$), only the phase of the condensate changes after passing the soliton-breather.

\section{Acknowledgments}
The author thanks the  participants of the S.P. Novikov' seminar for helpful discussions and Dr. Yifeng Mao for sending  the experimental data. Figures 1–3, taken from work \cite{Mao}, are published with permission of the American Physical Society.


\begin{thebibliography}{99}

\bibitem{Mao} Yifeng Mao,  Sathyanarayanan Chandramouli,  Wenqian Xu,  and Mark A Hoefer, 
{\it Observation of Traveling Breathers and Their Scattering in a Two-Fluid System},
 Physical Review Letters, {\bf 131}, 147201 (2023).
\bibitem{KuznetsovMikhailov} E.A. Kuznetsov, A.V. Mikhailov, {\it Stability of stationary waves in nonlinear media with weak dispersion}. Sov. Phys. JETP {\bf 40}, 855 (1975).
\bibitem{MaoHoefer} Y. Mao and M. A. Hoefer, {\it Experimental investigations of linear and nonlinear periodic travelling waves in a viscous fluid conduit}, J. Fluid Mech. {\bf 954}, A14 (2023).

\bibitem{ScottStevensonWhitehead} D. R. Scott, D. J. Stevenson, and J. A. Whitehead, {\it Observations of solitary waves in a viscously deformable
pipe}, Nature {\bf 319}, 759 (1986). 

\bibitem{OlsonChristensen} P. Olson and U. Christensen, {\it Solitary wave propagation
in a fluid conduit within a viscous matrix}, JGPR  {\bf 91}, 6367 (1986).


\bibitem{HelfrichWhitehead}K.R. Helfrich, J.A.  Whitehead , {\it Solitary waves on conduits of buoyant fluid in a more viscous
fluid}. Geophys. Astrophys. Fluid Dyn. {\bf 51}, 35–52 (1990).

\bibitem{magma}D.R. Scott, D.J. Stevenson, {\it Magma solitons}, GRL {\bf 11}, 1161-1164 (1984).

\bibitem{LowmanHoefer} N. K. Lowman and M. A. Hoefer, {\it Dispersive hydrodynamics
in viscous fluid conduits}, Phys. Rev. {\bf E 88},
023016 (2013).

\bibitem{kochina} P.Ya. Polubarinova-Kochina, {\it On the issue of shifting the oil-bearing contour}.   {\bf 47}, 254—257 (1945) (in Russian).

\bibitem{galin} L.A. Galin, {\it Nonstationary filtration with free boundaries}. Doklady AN SSSR {\bf 47}, 246-249 (1945) (in Russian).

\bibitem{SaffmanTaylor} P.G. Saffman , G. Taylor,   {\it The penetration of a fluid into a porous medium or Hele-Shaw cell
containing a more viscous liquid}, Proceedings of the Royal Society of London A: Math. Phys.  Engin. Sciences, {\bf 245} 312?329 (1958). 

\bibitem{Shabat} A.B. Shabat, 
{\it On the Korteweg-de Vries equation, 
Sov. Math., Dokl. {\bf 14}, 1266  (1973).

\bibitem{ZakharovShabat} V.E. Zakharov, A.B. Shabat, {\it Scheme of integration of nonlinear equations of mathematical physics by the method of inverse scattering problem}, Funct. Anal. Appl., {\bf 8}, 226  (1974).

\bibitem{WhittakerWatson} E.T. Whittaker and G. N. Watson, Modern Analysis,
Part 2, Cambridge University Press, 4th ed., 1962.


\bibitem{Novikov} S.P. Novikov, {\it Periodic problem for the Korteweg - de Vries equation}, Funct. Anal. Appl.
{\bf 8}, 54 -- 66 (1974). 

\bibitem{ZakharovManakovNovikovPitaevsky} V.E. Zakharov, S.V. Manakov, S.P. Novikov, L.P. Pitaevskii. {\it Theory of solitons: Method of inverse problem}, Moscow, Nauka, Fizmatlit 1980 (in Russian).

\bibitem{Kuznetso1977} E.A. Kuznetsov, 
 {\it Solitons in parametrically unstable plasma}, Sov. Phys. Dokl. {\bf 22}, 507  (1977).

\bibitem{KuznetsovSpectorFalkovich}
E.A. Kuznetsov, M.D. Spector, G.E. Falkovich, {\it On the Stability of Nonlinear Waves in Integrable Models}.
Physica {\bf 10D}, 379 (1984).
\bibitem{KuznetsovSpector} E.A. Kuznetsov, M.D. Spector, {\it Modulation instability of solitnon lattices in optical communcation systems},
Theor. Math. Phys. {\bf 120}, 997 (1999).

\bibitem{Kuznetsov-2017} E.A. Kuznetsov,  {\it  Fermi-Pasta-Ulam recurrence and modulation instability}, Pis'ma ZhETF {\bf 105}, 108 – 109 (2017) 
 [JETP Letters, {\bf 105}, 125-129 (2017)].

\bibitem{GardnerGreeneKruskalMiura} C. S. Gardner, J. M. Greene, M. D. Kruskal, and R. M. Miura, {\it Method for solving the Korteweg-de Vries equation}, Phys. Rev. Lett., {\bf 19}, 1095-1097 (1967).

 
\bibitem{ZakharovShabatNLS}  V.E. Zakharov, A.B. Shabat. {\it Exact theory of self-focusing and one-dimensional self-modulation of waves in nonlinear media}, Sov. Phys. JETP \textbf{34}, 62-69, (1972).

\bibitem{Krichever} I.M. Krichever, {\it Potentials with zeroth reflection coefficients  on the background of finite-zone ones }, Funct. Anal. Appl.,  {\bf 9}, 161 (1975).

\bibitem{TajiriWatanabe} M. Tajiri, and  Y. Watanabe, {\it Breather Solutions to the Focusing Nonlinear Schrodinger Equation}, Physical Review E {\bf 57}, 3510-3519 (1998). 

\bibitem{ZakharovGelash} V.E. Zakharov and  A.A. Gelash, {\it  Nonlinear Stage of Modulation Instability}. Physical Review Letters}, {\bf 111},  054101 (2013).




\end{thebibliography}
\end{document}